\documentstyle[osa,twocolumn]{revtex}   
          \def\dstyle{\displaystyle}
\def\hs{\hspace}                  \def\vs{\vspace}
\def\bc{\begin{center}}           \def\ec{\end{center}}
\def\beq{\begin{equation}}         \def\eeq{\end{equation}}
\def\bear{\begin{eqnarray}}       \def\eear{\end{eqnarray}}
\def\bt{\begin{tabular}}          \def\et{\end{tabular}}
\def\la{\langle}                  \def\ra{\rangle}
\def\lf{\left}                    \def\rt{\right}
\def\dg{\dagger}                  \def\ci{\cite}
\def\lb{\label}                   \def\ld{\ldots}
\def\pr{\prime}                   \def\sm{\small}

\def\td{\tilde}                   \def\pr{\prime}
\def\pd{\partial}                 
               
\def\nn{\nonumber}

\def\alf{\alpha}         
     \def\k{\kappa}
\def\Dlt{\Delta}     \def\dlt{\delta}    
\def\lam{\lambda}    \def\Lam{\Lambda}    \def\sig{\sigma}
\def\z{\zeta}      \def\vphi{\varphi}    \def\veps{\varepsilon}
\def\ome{\omega}      \def\Ome{\Omega}    
      
\begin{document}

\title{\hfill{\normalsize quant-ph/0012072 [JOSA A {\bf17} No 12 (2000)]}\\[10mm]
{\Large \bf Generalized uncertainty relations and coherent\\
      and squeezed states}}
\author{\bf D. A. Trifonov }
\address{Institute for Nuclear Research and Nuclear Energetics,
            72 Tzarigradsko Chauss\'ee, Sofia, Bulgaria}
\date{}
\maketitle

{\small
Characteristic uncertainty relations and their related squeezed states are
briefly reviewed and compared in accordance with the generalizations of
three equivalent definitions of the canonical coherent states.  The
standard SU(1,1) coherent states are shown to be the unique states that
minimize the Schr\"odinger uncertainty relation for every pair of the
three generators and the Robertson relation for the three generators. The
characteristic uncertainty inequalities are naturally extended to the case
of several states. It is shown that these inequalities can be written in
the equivalent complementary form. 

{\it OSIC codes}: 270.6570, 270.0270.}

\section{\sm\bf INTRODUCTION}

The uncertainty principle is a basic feature of quantum physics.  It was
introduced by Heisenberg \ci{H}, who demonstrated  an impossibility of the
simultaneous precise measurement of the coordinate ($q$) and momentum
($p$) canonical observables by positing an approximate relation $\Dlt p
\Dlt q \sim \hbar$, where $\hbar$ is the Planck constant.  This relation
was rigorously proved by Kennard \ci{K} in the form of the inequality
$(\Dlt p)^2(\Dlt q)^2 \geq \hbar^2/4$, where $(\Dlt X)^2$ is the variance
(dispersion) of the observable (Hermitian operator) $X$.  Robertson
\ci{R29} extended this inequality to arbitrary pair of operators $X$ and
$Y$,
\beq\lb{xyHUR}     
(\Dlt X)^2(\Dlt Y)^2 \geq \frac 14 \lf|\la[X,Y]\ra\rt|^2.
\eeq
The Heisenberg--Kennard--Robertson inequality \ci{H,K,R29} (\ref{xyHUR})
became known as the Heisenberg indeterminacy or uncertainty relation (UR)
for $X$ and $Y$, and here we shall follow this tradition.

According to inequality (\ref{xyHUR}) the product of the
uncertainties (precisions) $\Dlt X$ and $\Dlt Y$ of the measurements of
two quantum observables in one and the same state is not less than one
half of the absolute mean value of the associated observable $C=-i[X,Y]$.
Therefore this UR makes a statement about the preparation
of a quantum state.  However,  by using the technique of positive
operator-valued measure in measurement theory, one can extend UR
(\ref{xyHUR}) (with appropriately redefined notions of precisions $\Dlt X$
and $\Dlt Y$) to the joint measurement of two observables
in the form of a slightly more stringent inequality \ci{Ozawa}, with one
half instead of one fourth on the right-hand side of inequality 
(\ref{xyHUR}).

The UR's are formal expressions of the uncertainty principle in quantum
physics \ci{H,K,R29} and impose naturally fundamental limitations on the
accuracy of measurements and telecommunications. This problem became of
practical importance because of, e.g., experimental efforts to detect
gravitational waves \ci{Caves,Hollen}. The main problem is how to optimize
the intrinsic quantum fluctuations in the measurement process.
Significant progress has been achieved in this direction in the last two
decades by use the squeezed state technique \ci{Caves,Hollen}. The
concepts of squeezed state \ci{Caves,Hollen,Yuen} (SS) came from the
observation that the equality in the Heisenberg UR for the canonical
observables $p$ and $q$ can be maintained if the fluctuations of one of
the two observables are reduced at the expense of the other. So the UR's
play a dual role: They cause limitations on the measurement precision and
in the same time indicate ways to improve the accuracy of the measurement
devices. Thus a further study of the known UR's and their generalizations
is of both theoretical and practical importance.

In this paper recent developments in the field of generalized SS's and
UR's are considered and  some new results are reported. The concept of
SS's is closely related to that of coherent states \ci{KS,ZhangAli} (CS's)
introduced in 1963 patterned on the example of electromagnetic field
oscillators in the pioneering works by Glauber, Klauder and Sudarshan (see
Ref. 8, where a comprehensive list of references and reprints of selected
articles is provided). The SS's for the canonical observables (the
canonical SS's) are the unique one-mode states for which the three
definitions of the canonical CS's (Refs. 8 and 9) are equivalently
generalized, as is true in the multimode case as well \ci{T91}.

The paper is organized as follows.
The basic properties of the canonical CS's and canonical SS's are briefly
reviewed in Section 2. A new inequality is pointed out, the minimization
of which determines the canonical CS's uniquely.  In Section 3 we consider
the canonical SS's  and show that they can be defined in three equivalent
ways.  The generalization of the SS to the case of arbitrary two
observables on the basis of the more precise Schr\"odinger (or
Schr\"odinger--Robertson) UR \ci{SR} is considered in section 4.  SS's for
several observables on the basis of Robertson UR for $n$ observables are
discussed in Section 5. The extension of the Schr\"odinger--Robertson UR's
to all characteristic coefficients of the uncertainty matrix and to the
case of two and several states is the subject of Section 6. Some
applications of the ordinary and state-extended characteristic UR's are
outlined; the main applications are the construction of observable induced
metrics between quantum states  and the finer classification of states, in
particular of group--related CS's
\ci{KS}.

\section{\sm\bf THE CANONICAL CANONICAL COHERENT STATES}

The important overcomplete family $\{|\alf\ra\}$ of canonical CS's
$|\alf\ra$, $\alf \in {\mathbf C}$ (called also Glauber CS's), can be
defined in  three equivalent ways \ci{KS,ZhangAli}:\\[-2mm]

(D1) As the set of eigenstates of boson destruction operator (the
ladder operator) $a$:\,\, $a|\alf\ra = \alf|\alf\ra,$

(D2) As the orbit through the ground state $|0\ra$ ($a|0\ra = 0$)
constructed by use of the unitary displacement operators
$D(\alf)=\exp(\alf a^\dg - \alf^*a)$:\,\, $|\alf\ra = D(\alf)|0\ra$.

(D3) As the set of states which minimize the Heisenberg inequality
(\ref{xyHUR}) for the Hermitian components $p$ and $q$ of $a$ with equal
uncertainties $\Delta q = \Delta p$\,  ($a = (q+ip)/\sqrt{2}$; henceforth
we work with dimensionless observables).\\[-2mm]

Let us note that in the definition (D3) one requires the minimization of
inequality (\ref{xyHUR}) for $p,\,q$ plus the equality of the two
variances. The set of states which minimize inequality (\ref{xyHUR}) for
$p,\,q$ is much larger \ci{ZhangAli}. It is worth looking for another UR,
the minimization of which determines the CS's $|\alf\ra$ uniquely. Such UR
turned out to be the inequality
\beq\lb{new pqHUR}      
(\Dlt q)^2 + (\Delta p)^2 \,\geq\, 1,
\eeq
which follows from the obvious sequence
$(\Dlt q)^2 + (\Delta p)^2 \,\geq\, 2\Dlt p\Dlt q \,\geq\, 1$, and
therefore is less precise than the Heisenberg inequality  $\Dlt p\Dlt q
\geq 1/2$. 

The overcompleteness property reads ($\alf = \alf_1 + i\alf_2$,
d$^2\alf = {\rm d}\alf_1 {\rm d}\alf_2$)
\beq\lb{Res 1}          
1 = \int|\alf\ra\la\alf|d\mu(\alf),\quad {\rm d}\mu(\alf) = \frac 1\pi
{\rm d}^2\alf.
\eeq
One may say that the family $\{|\alf\ra\}$ resolves the unity operator
with respect to the measure $d\mu(\alf)$ (overcompleteness of
$\{|\alf\ra\}$ in the strong sense \ci{KS}).  This relation provides the
important analytic representation, known as canonical CS representation or
the Fock-Bargmann analytic representation, in which $a={\rm d}/{\rm
d}\alf,\,\, a^\dg = \alf$ and the state $|\Psi\ra$ is represented by the
function $\Psi(\alf) = \exp(|\alf|^2/2)\la\alf^*|\Psi\ra$. In the years
1963-64 Klauder (see the references in Ref. 8)  developed a general theory
of the continuous representations and suggested the possibility  of
constructing overcomplete sets of states by use of irreducible
representations of Lie groups.

There are at least three different ways (methods) of generalizing the
canonical CS's that correspond to definitions (D1)--(D3) above
\ci{ZhangAli}:\\[-2mm]

(D$^\pr$1). The diagonalization of a non-Hermitian operator $L\neq L^\dg$
(the eigenstate way, or the ladder operator method). The
corresponding overcomplete (in the weak or strong sense \ci{KS}) families
of states could be called $L$ CS's or ladder operator CS's. 

(D$^\pr$2).  The construction of orbit $\{|\vec{z}\ra\}$ through a fixed
vector $|\psi_0\ra$ of a family of unitary operators $D(\vec{z})$ (
orbit way or the displacement operator method). The corresponding
CS can be called $D$ CS's or displacement operator CS's.

(D$^\pr$3).  The minimization of appropriate UR $F[\psi] \geq 0$, where
$F[\psi]$ is a functional of states $|\psi\ra$ (the uncertainty
way). The corresponding overcomplete families of states could be called
$U$ CS's or (optimal) uncertainty CS's.\\[-2mm]

The first two methods, especially the second of these, have received 
considerable attention and have been videly
applied to various fields of physics \cite{KS,ZhangAli}, whereas the third
one received 
significant attention  only recently; see Refs. 12-25 and References
therein.  The developments of the second approach is thoroughly discussed
in Refs.  8 and 9. Therefore in Section 3 I 
provide a brief review of the main steps in the
first and the third ways only, noting main relationships between the three
general definitions $(D^\pr)$.  It appears that  the (multimode)
canonical SS's are the unique states, 
for which the three definitions $(D1) -- (D3)$ are
equivalently generalized.  It is worth noting that some authors (e.g.,
those of Ref. 9) were pessimistic about the possibility of effective and
useful generalization of the third defining property of canonical CS's to
the case of more-complicated systems.\\[3mm]

\section{\sm\bf THE CANONICAL SQUEEZED STATES AS $L$, $D$, and $U$ COHERENT
STATES} 

Canonical CS $|\alf\ra$ diagonalizes the operator $a$, $[a,a^\dg]=1$, which
is the ladder operator in the harmonic oscillator algebra $ho(1)$ spanned
by $\{1,a,a^\dg,a^\dg a\}$. The subalgebra spanned by $\{1,a,a^\dg\}$ is
known as the Heisenberg--Weyl algebra, $h(1)$.
This was the first and seminal example of
diagonalization of a non-Hermitian operator.  We stress that the
eigenstates of $a$ and of other non-Hermitian operators are not orthogonal;
the term "diagonalization" is used for brevity and in
analogy to the case of Hermitian operators.  Chronologically the second
example of $L$ CS, to the best of the author's knowledge, was given in
Refs. 26 and 27, where the diagonalization of the real and stationary \ci{MM}
and complex and time-dependent \ci{MMT} combinations of operators
$a,\,a^\dg$ has been performed ($\alf\in {\mathbf C}$),
\beq\lb{alf;t}       
\bt{c}$\dstyle 
A(t)|\alf;t\ra = \alf|\alf;t\ra,$\\
$\dstyle A(t) = u(t)a + v(t)a^\dg = A(u,v).$
\et
\eeq
The operator $A(t)$ was constructed \ci{MMT} as a non-Hermitian invariant
for the quantum varying frequency oscillator with Hamiltonian  $H =
(1/2)\left(p^2 + \ome^2(t)q^2/\ome_0^2\right)$,\,\, i.e., $A(t)$ had to
obey the equation $\pd A/\pd t - i[A,H] = 0$.  To satisfy this condition,
the parameter 
$\veps = (u-v)/\sqrt{\ome_0}$ was introduced and forced to obey the
classical oscillator equation $\ddot{\veps} + \ome^2(t)\veps = 0$. Here $H$
is also dimensionless, and $\ome_0$ is a frequency parameter that may be
taken as $\ome(0)$. The dimensional Hamiltonian is $\hbar\ome_0 H$.  The
boson commutation relation $[A,A^\dg]=1$ was  ensured by the Wronskian
$\veps^*\dot{\veps} - \veps\dot{\veps}^* =  2i$.  Then $\dot{\veps} =
i(u+v)\sqrt{\ome_0}$, $|u|^2 - |v|^2 = 1$, and $A(t) = U(t)A(0) U^\dg(t)$,
where $U(t)$ is the evolution operator and $A(0)=u_0a+v_0a^\dg$.

In the coordinate representation the wave functions take the form of an
exponential of a quadratic \ci{MMT} (for the reader's convenience in this
formula we restore the dimensions $x=q\sqrt{\hbar/m_0\ome_0}$,
$m_0$ being a mass parameter)
\bear\lb{<x|a,u,v>}        
\Psi_\alf(x,t)=\la x|\alf;t\ra =
\frac{(\pi l_0^2)^{-1/4}}{(u-v)^{1/2}}\qquad \nn \\
\times \exp\left[-\frac{1}{2l_0^2}\frac{v+u}{u-v} 
\left(x - \frac{\sqrt{2}\,l_0\alf}{u+v}\right)^2\right] \nn \\
\times\exp\lf[\frac{1}{2}\left(\frac{u^* + v^*}{u+v}\alf^2 - 
|\alf|^2\right)\rt],
\eear
where $l_0 = (\hbar/m_0\ome_0)^{1/2}$ (a length parameter).  Note that the
time dependence is embedded completely in $u(t)$ and $v(t)$ [or,
equivalently, in $\veps(t)$ and $\dot{\veps}(t)$] which justifies the
notation $|\alf;t\ra = |\alf,u(t),v(t)\ra$. The wave functions
[Eq. (\ref{<x|a,u,v>})] represent the time evolution of the canonical CS's
$|\alf\ra$ if the initial conditions \ci{MMT} $\veps(0) =
1/\sqrt{\ome_0},\,\, \dot{\veps}(0) = i\sqrt{\ome_0}$ are imposed [then
$u(0)=1,\, v(0)=0$]. Under these conditions $|\alf,u(t),v(t)\ra =
U(t)|\alf\ra$.  Time evolution of an initial $|\alf,u_0,v_0\ra$ for
quadratic Hamiltonian system was studied in Ref. 7, where eigenstates
of $ua + va^\dg$ were denoted as $|\alf\ra_g$. The invariant $A(t)$ in
Ref. 27 coincides with the boson operator $b(t)$ in Ref. 7. For
different purposes invariants and wave functions for one-dimensional
time-dependent quadratic systems were later studied in many papers
\ci{MM79,DodAng}.  Solutions to the Schr\"odinger equation for the
nonstationary systems have been previously obtained, e.g., by Husimi and
Chernikov \ci{HusCher}, but with no reference to the eigenvalue problem
and the invariants.  Gaussian wave functions such as Eq. (\ref{<x|a,u,v>})
have been studied by Schr\"odinger \ci{S2} and Kennard \ci{K}.

The nonstationary oscillator Hamiltonian is an element of the noncompact
algebra $su(1,1)$ in the representation with Bargmann indices $k=1/4,3/4$,
where the generators are ($K_\pm = K_1\pm iK_2$)
$$K_3 = a^\dg a/2 + 1/4,\quad K_- = a^2/2,\quad  K_+ = {a^\dg}^2/2.$$
Therefore $U(t)\in SU(1,1)$  and the set $\{|\alf,u(t),v(t)\ra\}$ is an
 $SU(1,1)$ orbit through the initial CS $|\alf;u_0,v_0\ra$. At $u_0=1$,
$v_0=0$ this is an orbit through $|\alf\ra = D(\alf)|0\ra$,
\beq\lb{UD|0>}           
|\alf,u(t),v(t)\ra = U(t)|\alf\ra = U(t)D(\alf)|0\ra.
\eeq
The parameters $u$ and $v$ refer to the SU(1,1), and $\alf$ refers to the
Heisenberg--Weyl group H(1). The operator $U(t)D(\alf)$ belongs to the
semidirect product SU(1,1)$\wedge$H(1); therefore  the set of states
$|\alf,u,v\ra$ is an orbit of the group SU(1,1)$\wedge$H(1)
\ci{T91,Nagel95}.  This establishes the equivalence of the first two
definitions (D$^\pr$1) and (D$^\pr$2) for the states $|\alf,u,v\ra$.

For non-quadratic Hamiltonians the invariant (the new boson
annihilation operator) $A(t)=U(t)A(0)U^\dg(t)$ is not linear in $a$ and
$a^\dg$ and its eigenstates are no more of the form $|\alf,u(t),v(t)\ra$
\ci{M72,T91}. Therefore the term "coherent states for the
nonstationary oscillator"  \ci{MMT} for $|\alf;t\ra = |\alf,u,v\ra$  is
indeed adequate. The Hermitian components $P(t)$ and $Q(t)$ of $A(t)$ are
also invariant, obey the canonical commutation relations $[Q(t),P(t)]=i$,
and have the physical meaning of coordinates of the initial point in the
phase space \ci{M72}. Nonlinear realizations of boson operators
$A_{(k)}$ ($k$-photon operators) are considered in the first paper of
Ref. 35.

The orthonormalized eigenstates $|n,u,v\ra$ of the quadratic invariant
$A^\dg(t)A(t)$ (an element of su(1,1)) were also constructed in
Ref. 27.  Note that any power of $A(t)$ and $A^\dg(t)$ is again an
invariant. In particular $A^\dg(t)A(t)$ coincides with the
Ermakov--Lewis invariant \ci{EL}.  At the appropriate initial
conditions the eigenstates of $A^\dg(t)A(t)$ represent the time evolution
of the Fock states $|n\ra$.

For the $s$-dimensional quadratic system there are $s$ linear in $a_\mu$
and $a^\dg_\mu$ invariants $A_\mu(t) = u_{\mu\nu}(t)a_\nu +
v_{\mu\nu}(t)a^\dg_\nu \equiv A_\mu(u,v)$, which were simultaneously
diagonalized \ci{HMMT},
\beq\lb{A_mu|a,u,v>}      
A_\mu(u,v)|\vec{\alf},u,v\ra = \alf_\mu|\vec{\alf},u,v\ra,
\quad \mu = 1,\ld, s.
\eeq
The wave functions $\la\vec{x}|\vec{\alf},u,v\ra$ are $s$ dimensional
Gaussians. This set of states is an orbit of the group Sp$(s,R)\wedge$
H$(s)$ through the $s$ dimensional vacuum state $|0\ra$. In 
Section 4 we shall see that it can be considered family of $U$ CS's
related to the Robertson UR for $2s$ canonical observables \ci{R34}.
Invariants and wave functions for nonstationary $s$-dimensional quadratic
systems were later studied by many authors (see Ref. 28 and references
therein). 

By means of the known BCH formula for the transformation $S(\z)aS^\dg(\z)
= \cosh|\z|\,a - \sinh|\z|\exp(i\vphi)\,a^\dg$, $\z=|\z|\exp(i\vphi)$,
with the operator
$S(\z) = \exp[(\z a^{\dg 2} - \z^*a^2)/2] =  \exp(\z K_+ - \z^*K_-)$
the solutions $|\alf,u,v\ra$ for the one-dimensional oscillator systems
are immediately brought, up to a phase
factor, to the form of famous Stoler states $|\z,\alf\ra = S(\z)|\alf\ra$
(Ref. 39) with ${\rm cosh}|\z| = |u|,\,\, \vphi
= {\rm arg}\,v - {\rm arg}\,u$,
\beq\lb{Stolform}        
|\alf,u,v\ra = \exp(i{\rm arg}u)\,\exp(\z K_+ - \z^*K_-) |\alf\ra.
\eeq

Yuen \ci{Yuen} called the eigenstates $|\alf,u,v\ra$ of $ua+va^\dg$
\, two photon CS's\, and suggested that the output radiation of an ideal
monochromatic two photon laser is in a state $|\alf,u,v\ra$. Hollenhorst
\ci{Caves} named these states \, squeezed states\,  to reflect that they
 exhibit fewer fluctuations in $q$ or $p$ than those in CS
$|\alf\ra$. It is convenient to call these SS's canonical. They were
intensively studied in quantum optics and are experimentally realized (see
references in Ref. 40).  The eigenstates $|n,u,v\ra$ of
$(ua+va^\dg)^\dg(ua+va^\dg)$ became known as squeezed Fock states
($|n\!=\!0,u,v\ra$ -- squeezed vacuum) and the operator $S(\z)$ 
as a (canonical) squeeze operator \ci{LK,ZhangAli}.  Eigenstates
$|\vec{\alf},u,v\ra$ [Eq.  (\ref{A_mu|a,u,v>})] are known as multimode
(canonical) SS's \ci{nmodeSS}.

Radcliffe and Arecchi {\it et al} \ci{RadGil} introduced and studied the
SU(2) 
analog of the states $|\alf\!=\!0,u,v\ra$ in the similar form to that of
Stoler states [Eq (\ref{Stolform})] i.e., spin CS's or atomic CS's.
The results of Ref. 42 about the SU(2) CS's have been extended by Perelomov
\ci{Perel} to the
noncompact group SU(1,1) and to any Lie group $G$ as well; he succeeded in
proving the Klauder suggestion for construction
of overcomplete families of states by using unitary irreducible
representations of Lie groups [group-related CS's (Ref. 8)].  For the
discrete series $D^{(+)}(k)$ of SU(1,1), $k=1/2,1,\ld$, and the 
lowest-weight reference vector these CS's [CS's with maximal symmetry or
the standard SU(1,1) CS's] take a form similar to Eq. (\ref{Stolform}):
\bear\lb{|xi;k>}         
|\xi;k\ra &=&
\exp(\z K_+ - \z^* K_-)|k,k\ra \nn \\
          &=& (1-|\xi|^2)^k\,\exp(\xi K_+) |k,k\ra,
\eear
where $|\xi|= {\rm tanh}|\z|\in {\mathbf D}_1$, arg$\xi = -{\rm arg}\z + \pi$.
The relation of these CS's to the $U$ CS's, definition (D$^\pr$3), was
established later \ci{T94,T97} on the basis of Schr\"odinger and Robertson
UR's (see Section 4 below).\\

The third and seminal example of diagonalization of non-Hermitian operator
was given in 1971 by Barut and Girardello \ci{BG}, where they constructed
the eigenstates of the SU(1,1) ladder operator $K_-$ in
the discrete series $D^{(\pm)}(k)$ and proved the overcompleteness of its
eigenstates $|z;k\ra$,
\beq\lb{BGCS}           
|z;k\ra = N_{BG}\sum_{n = 0}^{\infty}
\frac{z^{n}}{\left(n!\Gamma(2k+n)\right)^{1/2}} |k,k+n\rangle,
\eeq
where $N_{BG} =  [\Gamma(2k)/{}_0F_1(2k;|z|^2)]^{\frac{1}{2}}$, and
$_0F_1(c;z)$ is the confluent hypergeometric function.  The family of
Barut--Girardello (BG) CS's $|z;k\ra$ resolves the unity operator and
provides a new analytic representation \ci{BG}, which has been used in the
diagonalization of more general su$^c$(1,1) operators
\ci{T94,T96,Brif97,T97,T98b}. 
This representation was recently extended to the boson realizations of the
higher dimensional algebras $u(N,1)$ \ci{FujFun} and $u(p,q)$ \ci{T98a}.

For further developments in the direction of $L$- CS's, including the cases
of Weyl ladder operators for the $q$-deformed h$_q$, su$_q$(2) and
su$_q$(1,1), see, e.g., the brief review in Ref. 47 and references therein. 
The nonlinear CS's \ci{MMSZ}, which have enjoyed increasing interest 
recently \ci{nlcs}, are also defined as eigenstates of non-Hermitian
operators.

Every set of eigenstates $|z\ra$ of a fixed non-Hermitian operator $L$,
$L|z\ra = z|z\ra$, in particular the set of nonlinear CS's \ci{MMSZ,nlcs},
the BG CS's $|z;k\ra$ and their $q$--deformed extension, can be
also defined according to (D$^\pr$3) with equal
variances of the Hermitian components $X$ and $Y$ of $L$, $L = X + iY$, on
the basis of UR (\ref{xyHUR}). The requirement of equal variances may be
omitted if one finds a suitable less-precise inequality. It turned out that
these same $L$ CS can be uniquely defined as states which minimize the
less-precise UR
\beq\lb{new xyHUR}
(\Dlt X)^2 + (\Dlt Y)^2 \geq \lf|\la[X,Y]\ra\rt|
\eeq
for the components of $L$. The proof of inequality (\ref{new xyHUR})
consists in the 
observation that $(\Dlt X)^2 + (\Dlt Y)^2 \geq 2\Dlt X \Dlt Y \geq
\lf|\la[X,Y]\ra\rt|$. The minimization of inequality (\ref{new xyHUR})
occurs in the 
states with equal $\Dlt X$ and $\Dlt Y$ only: $(\Dlt X)^2 = (\Dlt Y)^2 =
\lf|\la[X,Y]\ra\rt|/2$.
In general the representation of $L$ CS's as $D$ CS's is not possible, as 
proved for the family of BG CS's $|z;k\ra$ \ci{T98b}.

The larger family of canonical SS's $|\alf,u,v\ra$ can be uniquely
determined \ci{DKM,T91} in the third equivalent way (as $U$ CS) on the
basis of the more-precise Schr\"odinger UR \ci{SR} for $p$ and $q$,

\beq\lb{pqSUR}         
(\Dlt p)^2 (\Dlt q)^2 - (\Dlt pq)^2 \,\geq\, 1/4,
\eeq
where $\Dlt pq$ is the covariance of $p$ and $q$, $\Dlt pq = \la
pq+qp\ra/2 - \la p\ra\la q\ra$.
The three second moments of $p$ and $q$ in $|\alf,u,v\ra$ do not depend on
$\alf$ and read as \ci{DKM} $(\Dlt q)^2 = \frac 12|u-v|^2$,\, $(\Dlt p)^2 =
\frac 12|u+v|^2$,\, $\Dlt pq = -{\rm Im}(uv^*)$.  In other parameters they
were calculated by Kennard \ci{K}, Stoler \ci{Stol}, in Ref. 10, and in
Ref. 34. The above moments saturate inequality (\ref{pqSUR})
identically with respect to $u,\,v$. One sees that the variance of $p$
($q$) tends to zero when $v\rightarrow -u$ ($v\rightarrow u$). Therefore
these states can be called $q$--$p$ ideal SS's.

By construction, the set $\{|\alf,u,v\ra\}$ is stable under the action of
the evolution operator $U(t)$ of the varying frequency oscillator,
$U(t)|\alf,u_0,v_0\ra = |\alf,u(t),v(t)\ra$. It was shown \ci{T91} that
the most general Hamiltonian that keeps the canonical SS's stable is
quadratic in $p$ and $q$. If the time evolution is governed by a
time-dependent quadratic Hamiltonian
$H(t) = g_1(t)p^2 +g_2(t)(pq+qp) + g_3(t)q^2$ [where $g_i(t)$ are arbitrary
differentiable functions] then an
initial wave function of the form of Eq. (\ref{<x|a,u,v>}), an initial SS,
keeps 
this form for later times with some time-dependent $u(t)$ and $v(t)$.
Here again the time dependence of the wave function $\Psi_\alf(x,t)$
is completely embedded into parameters $u(t)$ and $v(t)$. $u(t)$ and
$v(t)$ can be expressed in terms of $g_i(t)$ and a classical function
$\veps(t)$,
which obeys the  oscillator equation $\ddot{\veps} +\Ome^2(t)\veps
= 0$ with "frequency" \ci{HMMT} $\Ome^2(t) = 4\ome^2_0g_1g_3
+2\ome_0g_2\dot{g_1}/g_1 + \ddot{g}_1/2g_1 - 3\dot{g}_1^2/4g_1^2 -
4\ome^2_0 g_2^2 - 2\ome_0\dot{g}_2$ and Wronskian $\veps^*\dot{\veps} -
\veps\dot{\veps}^* = 2i$. Here $g_i(t)$ are dimensionless, $[\veps] =
[\ome_0]^{-1/2}$.

It is worth noting that the essential state parameters of SS
$|\alf,u,v\ra$ (up to a phase factor) are four: In view of $u\neq 0$ one
can rescale the parameters in Eq. (\ref{alf;t}) by dividing both sides by $u$.
These parameters can be chosen in the form of two canonically  conjugated
pairs of classical observables \ci{T91}:
$\la p\ra,\,\,\la q\ra$ and ${\td p} = \Dlt pq/\Dlt q,\,\, {\td q} =
\Dlt q$.  For quadratic systems they satisfy the classical equations
with Hamiltonian function ${\cal H} = \la v(t),u(t),\alf|
H|\alf,u(t),v(t)\ra = {\cal H}(\la p\ra,\la q\ra,{\td p},{\td q})$,
\bear          
\frac{d\la p\ra}{dt} = - \frac {\pd{\cal H}}{\pd\la q\ra},\qquad
\frac{d\la q\ra}{dt}= \frac{\pd{\cal H}}{\pd\la p\ra}, \lb{classeqs2a}\\[3mm]
\frac{d{\td p}}{dt} = - \frac{\pd{\cal H}}{\pd {\td q}},\qquad
\frac{d{\td q}}{dt} = \frac{\pd{\cal H}}{\pd {\td p}}.\lb{classeqs2b}  
\eear
The stable evolution of quantum SS's is governed by these classical
canonical equations. The time evolution of squeezing is controlled by the
classical eqs. (\ref{classeqs2b}).
If one restores the dimensions, one finds that in the limit
$\hbar = 0$ the noisy variables ${\td p}$ and ${\td q}$ vanish, whereas the
Eq. (\ref{classeqs2a}) recover the canonical classical equations
with quadratic Hamiltonian function.

A sequence of different subsets of $\{|\alf,u,v\ra\}$ can be determined
uniquely from the sequence of the UR's considered above: $[(\Dlt p)^2 +
(\Dlt q)^2]^2/4 \geq (\Dlt p)^2 (\Dlt q)^2 \geq (\Dlt p)^2 (\Dlt q)^2 -
(\Dlt pq)^2 \geq 1/4$.

Thus the family of canonical SS's can be regarded equivalently as
$L$ CS's \,($L= ua+va^\dg$),\,\, $D$ CS's \,($D = \exp[(\z a^{\dg 2} - \z
a^2)/2]\,\exp(\alf a^\dg - \alf^*a)$),\, and\, $U$ CS's \,(Schr\"odinger $U$
CS, Schr\"odinger optimal uncertainty states, correlated states \ci{DKM}
or Schr\"odinger intelligent states \ci{T94,T96}, the term "intelligent
states" being introduced in Ref. 50).

\section{\sm\bf SCHR\"ODINGER INEQUALITY AND \\SQUEEZED 
STATES FOR TWO GENERAL \\ OBSERVABLES} 

The concept of SS has been extended to noncanonical pair of observables,
in particular to two generators of an arbitrary Lie group \ci{Wodki}
on the basis of the equality in the Heisenberg UR: A
set of SS's for two observables (Hermitian operators) $X$ and $Y$ was
defined as the set of solutions to the eigenvalue equation $(X+i\lam
Y)|z,\lam\ra = z|z,\lam\ra$, where $\lam$ is real parameter.  Solutions to
this equation for $X$ and $Y$, the quadratures of $a^2$, 
were constructed in Ref. 14.  A criterion was proposed \ci{Wodki} according to
which a state $|\psi\ra$ is squeezed if $(\Dlt X)^2$ or $(\Dlt Y)^2$ is
less than $|\la[X,Y]\ra|/2$. This construction was
generalized and refined in Ref. 15. The points are that the equality in
inequality (\ref{xyHUR}) is not invariant under the linear transformations
of $X$ and 
$Y$, in particular under the linear canonical transformations \ci{SCB95}
and the inequality $(\Dlt X)^2 \leq |\la[X,Y]\ra|/2$ can hold \ci{T94,T97}
for very large values of the fluctuation $(\Dlt X)^2$.  For example, the
standard SU(1,1) CS $|\xi;k\ra$ can exhibit strong squeezing according to
the Eberly-Wodkiewicz criterion, whereas the
fluctuations $(\Dlt K_1)^2$ and $(\Dlt K_2)^2$ are always grater than or
equal to their value of $k/2$ in the ground state $|k,k\ra$ \ci{T94}.
Besides, the Heisenberg UR for $K_1$ and $K_2$ is not minimized in every
CS $|\xi;k\ra$.

The appropreate UR to be used for the definition of
SS's for two general observables $X$ and $Y$ is that of Schr\"odinger (or
Schr\"odinger--Robertson) \ci{SR},
\beq\lb{xySUR}
(\Dlt X)^2 (\Dlt Y)^2 \geq \frac 14\left|\la[X,Y]\ra\right|^2 +
(\Dlt XY)^2,
\eeq
where $\Dlt XY \equiv \la XY+YX\ra/2 -\la X\ra \la Y\ra$ is the covariance
of $X$ and $Y$. It is more precise than inequality (\ref{xyHUR}) and is
reduced to  that inequality when $\Dlt XY=0$. The set of $X$--$Y$ SS's was
defined \ci{T94} as the set of states that minimize inequality
(\ref{xySUR}).  Such minimizing states were called generalized intelligent
states in Ref. 15. They could also be called $X$--$Y$ correlated states
\ci{DKM} or (Schr\"odinger) optimal uncertainty states (optimal US). It
was established that the $X$--$Y$ SS's can be defined as solutions to the
equations $(\lam X+iY)|\psi\ra = z|\psi\ra$, where $\lam$ is complex
parameter. To include the eigenstates of $X$, when they exist, one has to
relax this condition slightly:
\beq\lb{SOUS}
[u(X-iY) + v (X+ iY)]\,|z,u,v\ra = z|z,u,v\ra,
\eeq
$u,\,v,\,z \in {\mathbf C}$.
The three second moments of $X$ and $Y$ in solutions
$|z,u,v\ra$ read as  
\bear\lb{SOUSmeans} 
\bt{c}
$\dstyle
(\Dlt X)^2 = \frac 12 \frac{|u-v|^2}{|u|^2-|v|^2}\,i\la[X,Y]\ra,$\\
$\dstyle
\Dlt XY = \frac{{\rm Im}(u^*v)}{|u|^2-|v|^2}\,i\la[X,Y]\ra.$\\
$\dstyle (\Dlt Y)^2 = \frac 12\frac{|u+v|^2}{|u|^2-|v|^2}\,
i\la[X,Y]\ra.$ \et
\eear
It turned out that the standard SU(1,1) and SU(2) CS's also minimize the
Schr\"odinger inequality for the generators $K_1,\,K_2$ and $J_1,\,J_2$
respectively \ci{T94}. For example, the SU(1,1) CS's $|\xi;k\ra$ are
 a particular case of the $K_1$--$K_2$ optimal US $|z,u,v;k\ra$ 
corresponding to $z=-2k\sqrt{-uv}$ and $\xi = (-v/u)^{1/2}$.

The optimal US $|z,u,v\ra$ can exhibit arbitrary strong squeezing of $X$
and $Y$ when the parameter $v$ tends to $\pm u$\, \ci{T94}.  Therefore the
family of $|z,u,v\ra$ is a family of  $X$--$Y$ ideal SS's.  It is
worth noting an important application of the $K_i$--$K_j$ and $J_i$--$J_j$
optimal US in the quantum interferometry: As shown by Brif and Mann the
SU(1,1) and SU(2) intelligent states $|z,u,v\ra$ which are not
group-related CS's can greatly improve  the sensitivity of the SU(2) and
SU(1,1) interferometers \ci{BrifMann}.  Schemes for generation of
SU(1,1) and SU(2) optimal US of radiation field can be found, e.g., 
in Refs. 23, 51, and 52.

From $(\Dlt X)^2 \geq 0$ and Eqs. (\ref{SOUSmeans}) it follows that if the
commutator $i[X,Y]$ is positive (negative) definite then normalized
eigenstates of $u(X-iY) + v (X + iY)$ exist for $|u|>|v|$ ($|u|<|v|$) only
\ci{T97}. In such cases one can rescale the parameters and put
$|u|^2-|v|^2 =1$ ($|u|^2-|v|^2=-1$) as one normally does in the canonical
case.  It is also seen from Eqs. (\ref{SOUSmeans}) that in the states with
large $|\la[X,Y]\ra|$ the variances of both $X$ and $Y$ can be large.
Thus the frequently used term "minimum uncertainty states" for states that
minimize inequality (\ref{xyHUR}) or (\ref{xySUR}) is in fact adequate  in
the case of canonical observables only: In general the lowest level,
$\Dlt_0 \leq \Dlt X = \Dlt Y$, can be reached in some subsets of
$|z,u,v\ra$. It then is natural for a given state to be considered
squeezed if $\Dlt X$ or $\Dlt Y$ is less than $\Dlt_0$ \ci{T94,T98b}.

The family of Schr\"odinger optimal US for the two  quasi-spin operators
$K_1$ and $K_2$ in the series $D^+(k)$ was first constructed, up to a
normalization factor, in Ref. 15, and for the spin
operators $J_1$ and $J_2$ in Ref. 50  (with no
reference to Schr\"odinger inequality).  For the quadratures of $a^2$ the
states $|z,u,v\ra$ were constructed in Ref.  20 and in fact Ref. 15.  The
even and odd CS's \ci{DMM}, which are the first examples of the
intensively discussed macroscopic superpositions (see, e.g., Refs.  46,
49, and 54 and references therein), saturate inequality (\ref{xySUR}) with
vanishing covariance because they are eigenstates of $a^2$.  Minimization
of inequality (\ref{xySUR}) for the quadratures of the $q$-deformed boson
operator $a_q$ is considered in Ref. 55, and for the quadratures of the
$q$-deformed su(1,1) ladder operator $K_-(q)$ in Ref.  47.

Finally in this section, let us note that the minimization of the three
UR's with increasing precision, inequalities (\ref{new xyHUR}),
(\ref{xyHUR}) and (\ref{xySUR}),
\bear
\frac 14[(\Dlt X)^2 + (\Dlt Y)^2]^2 \,\geq\, (\Dlt X)^2 (\Dlt Y)^2 
\,\geq\, \nn \\
(\Dlt X)^2 (\Dlt Y)^2 - (\Dlt XY)^2 \,\geq\,
\frac 14\left|\la[X,Y]\ra\right|^2, \nn
\eear
determines naturally a sequence of subsets of $\{|z,u,v\ra\}$. One has
\beq\lb{setsequence}     
\{|z\ra\} \,\subset\, \{|z,u,v\ra|_{{\rm Im}uv^*=0}\}
\,\subset\, \{|z,u,v\ra\},
\eeq
where the smallest subset $\{|z\ra\}$ consists of states $|z\ra$ that
minimize the less precise UR [inequality (\ref{new xyHUR})].  These are
solutions to any of the two eigenvalue equations $(X\pm iY)|z\ra =
z|z\ra$, and, in the case of $X=q$, $Y=-p$\, ($X=K_1$, $Y=K_2$),\,
coincide with the Glauber CS's\, (BG CS).  They are the eigenstates
$|z\ra$  that are the extension of the Glauber CS's to any two
observables; The nontrivial example is given by the BG CS's $|z;k\ra$.
However the set of $|z\ra$ is not always continuous.  Nevertheless, this
set of  eigenvectors may still be "complete" in the finite Hilbert space,
as one readily sees in the case of spin $1/2$, for example.

\section{\sm\bf ROBERTSON INEQUALITY AND SQUEEZED STATES FOR \mbox{\large$n$} 
OBSERVABLES}

Two Hermitian operators (two observables) never close an algebra. Even in
the simplest case of Heisenberg--Weyl algebra h(1) in fact the operators
are three: $p$, $q$ and $1$ (the identity).
Physical systems with higher symmetry are described by three and more
independent observables. It was Robertson \ci{R34} who first realized that
there must be an UR "for all observables under consideration", "for we
cannot in general expect that the conditions necessary to insure minimum
uncertainty in one pair will be consistent with those which insure the
minimum in other pairs".
The relation that Robertson found for $n$ observables $X_1,X_2,\ld X_n$ 
reads as
\beq\lb{RUR}
{\rm det}\,\sigma(\vec{X}) \,\geq\, {\rm det}\, C(\vec{X}),
\eeq
where $\vec{X} \equiv (X_1,X_2,\ld,X_n)$, $\sigma(\vec{X})$ is the
$n\times n$ matrix of the second moments of observables, $\sigma_{ij} =
\la X_iX_j+X_jX_i\ra/2 - \la X_i\ra\la X_j\ra \equiv
\Dlt X_i X_j$, $i,j = 1,2,\ld, n$, and $C(\vec{X})$ is the $n\times n$
matrix of the first moments of the commutators $[X_i,X_j]$, $C_{kj} =
-\frac i2\la[X_k,X_j]\ra$.  For $n=2$, inequality (\ref{RUR}) coincides
with (\ref{xySUR}).  With minor changes the Robertson proof of inequality
(\ref{RUR}) is provided in the appendix of Ref. 47.

The Schr\"odinger UR proved to be efficient in defining the ideal SS for
two observables \ci{T94}, which encouraged us to define the SS's for
several general observables as states that minimize Robertson UR
(\ref{RUR}) \ci{T98b}. The latter definition is more effective for the
even number of observables, $n=2s$, because for odd
number the right-hand side of UR (\ref{RUR}) is vanishing.  The
minimization of UR (\ref{RUR}) is considered in detail in Ref. 22, 
where the minimizing states are called
Robertson intelligent states \ci{T97} or Robertson optimal US's \ci{T99}.
For even-number $n=2s$ the minimizing states are
eigenstates of one real or $s$ complex linear combinations of $X_j$, 
whereas for odd $n$ the diagonalization of one real
combination is necessary and sufficient.

For even $n=2s$, keeping the analogy to the case of two observables, we
define the $s$ "ladder" operators $\tilde{a}_\mu = X_\mu +iX_{\mu+s}$ and
their $s$ complex combinations
\beq
{\td A}_\mu(u,v) := u_{\mu\nu}\td{a}_\nu +
v_{\mu\nu}\tilde{a}^\dg_\nu = \beta_{\mu j}X_j,
\eeq
where $\beta_{\mu\nu} = u_{\mu\nu}+v_{\mu\nu}$, $\beta_{\mu,s+\nu} =
i(u_{\mu\nu} - v_{\mu\nu})$.
The condition for the equality in (\ref{RUR}) is the simultaneous
diagonalization of ${\tilde A}_\mu(u,v)$:
\beq\lb{|vec{z},u,v>}   
{\tilde A}_\mu(u,v)|\vec{z},u,v\ra = z_\mu|\vec{z},u,v\ra,
\eeq
$\mu =1,\ld,s$. In the minimizing states $|\vec{z},u,v\ra$ the second
moments of $X_\mu$ can be expressed in terms of the first moments of their
commutators:
\bear\lb{ROUSmeans}   
\sig = {\cal B}^{-1}\left( \bt{cc} $0$ & $\td{C}$\\
$\td{C}^{\rm T}$ & $0$ \et\right) {\cal B}^{-1}{}^{\rm T}, \\
{\cal B} =  \left(\bt{cc} $u+v$&$i(u-v)$\\
$u^* + v^*$&$i(v^* - u^*)$\et\right),\nn
\eear
where $\td{C}_{\mu\nu} = \frac 12\la [\td{A}_\mu,\td{A}_\nu^\dg]\ra$ and
$\sig = \sig(\vec{X};z,u,v)$ is the dispersion matrix.  Note that here
$u,\,v$ and $\td{C}$ are $s\times s$ matrices, $\beta$ is an $s\times n$
matrix,  and ${\cal B}$ and $\sig$ are $n\times n$, $n=2s$.  We suppose
that ${\cal B}$ is not singular. For two observables, $n=2$, we have
$\beta_{11} = u+v$, $\beta_{12}= i(u-v)$,  and $\td{C} = i(|u|^2 -
|v|^2)[X_1,X_2]$, and Eqs. (\ref{ROUSmeans}) recover Eqs.
(\ref{SOUSmeans}).  From Eq. (\ref{|vec{z},u,v>}) and the equivalence
$\Dlt X\,(\psi) = 0\quad \leftrightarrow\quad X|\psi\ra = x|\psi\ra$ it
follows that the variance $\Dlt X_i (\vec{z},u,v)$ will tend to zero when
$\beta_{\mu,k}\rightarrow 0$ for every $k\neq i$ and at least for one
$\mu$. If this can be managed for every $i=1,\ld,n$ then the set of
$|\vec{z},u,v\ra$ is a set of ideal SS for $n$ observables.\\

{\bf A. Example 1: The canonical observables} \\
Let $X_\mu = q_\mu,\,\, X_{s+\mu} = p_\mu$, where
$p_\mu,\,q_\mu$ are $s$ pairs of canonical observables,
$[q_\mu,p_\nu]=i\dlt_{\mu,\nu}$. In this case $\td{a}_\mu = a_\mu\sqrt{2}$
and $\td{A}_\mu = A_\mu(u,v)\sqrt{2}$, where $A_\mu(u,v)$ are the
Bogolyubov transforms of boson creation-annihilation operators. Their
common eigenstates $|\vec{\alf},u,v\ra$ [Eq. (\ref{A_mu|a,u,v>})], were
constructed in Ref. 37 and studied in many papers \ci{nmodeSS} (but with
no reference to the Robertson relation). Up to phase factors they coincide
with the multimode (canonical) SS's \ci{nmodeSS}. We note here that they
are ideal SS's for all $p_\mu$ and $q_\mu$. The proof is based on the fact
that the uncertainty matrix $\sig(\vec{Q})$, $\vec{Q} = (\vec{p},\vec{q})$
is nonnegative definite and can be diagonalized by means of linear
canonical transformations \ci{SCB95,T95}: $\vec{Q}\,\rightarrow\,
\vec{Q}\,^\pr = \Lam \vec{Q}$.  The total symplectic $\Lam$ preserves the
equality in the Robertson relation \ci{T97}. The variances of the
$\Lam$-transformed operators in the old state are equal to that of the old
operators in the new state, $|\psi^\pr\ra = U(\Lam)|\psi\ra$, where $U$ is
the unitary generator of the symplectic transformation. So the new state
is also Robertson optimal US if the old one is and vice versa. However,
$|\vec{\alf},u,v\ra = U(u,v)D(\vec{\alf})|0\ra$, where the operator
$U(u,v)D(\vec{\alf})$ belongs to the semidirect product group
Sp$(s,R)\wedge$H$(s)$. Thus the canonical multimode SS's are
simultaneously $L$, $D$, and $U$ CS.

Becuase of the canonical commutation relations, the
general expression [Eqs. (\ref{ROUSmeans})] for the uncertainty matrix
simplifies: In states $|\vec{\alf},u,v\ra$ the $s\times s$ matrix $\td{C}$
is a multiple of the identity, $\td{C} = 1/2$. As a result $\sig(\vec{Q})$
becomes symplectic itself.  In other parameters this $\sig(\vec{Q})$ was
calculated by Ma and Rhodes \ci{nmodeSS}.

The macroscopic superpositions $|\vec{\alf}\ra_\pm $ of multimode Glauber
CS's $|\vec{\alf}\ra$ and $|\!-\!\vec{\alf}\ra$ [the even and odd
multimode CS's (Refs. 56)] are eigenstates of $a_\mu a_\nu$.
Therefore $|\vec{\alf}\ra_\pm $ minimize Robertson UR for the quadratures
of $a_\mu a_\nu$. Note that $a_\mu a_\nu$ are mutually commuting Weyl
ladder operators of the algebra sp$(n,R)$. Therefore $|\vec{\alf}\ra_\pm $
are the BG-type CS's for $sp(s,R)$ \ci{T98a}.\\

{\bf B. Example 2: The three generators $K_i$ of SU$(1,1)$.}\\
For odd number of observables the Robertson UR is minimized in the
eigenstates of their real combinations only. In this case the more general
set of eigenstates $|z,u,v,w;k\ra$ of complex combination $uK_- + vK_+ +
wK_3$ of $K_i$ (the general operator of $su^c(1,1)$) was constructed
\ci{T96,Brif97}. In terms of the orthonormalized eigenstates $|k,k+n\ra$
of $K_3$ the states with $u\neq 0$ read
\bear
|z,u,v,w;k\ra
= {\cal N} \sum_{n=0}^{\infty}\left(-\frac{l+w}{2u}\right)^n
\sqrt{\frac{(2\k)_n}{n!}}\nn \\
\times\,_2F_1\left(\k+\frac{z}{l},-n;2\k;\frac{2l}{l+w}\right)\,
|k,n+k\ra, \lb{|zuvw>}
\eear
where ${\cal N}$ is the normalization factor (the explicit form ${\cal
N}(z,u,v,w,k)$ can be found in \ci{T98b}),\, $l = \sqrt{w^2-4uv}$, $(a)_n$
is Pochhammer symbol and $_2F_1(a,b;c;z)$ is the Gauss hypergeometric
function. The states are normalizable if at least one of the two
inequalities $|w\pm\sqrt{w^2-4uv}| <  2|u|$ holds. These states minimize
Robertson UR (\ref{RUR}) for the three operators $K_i$ iff Im$w=0$,
$v=u^*$. Among the minimizing states there are the standard $SU(1,1)$ CS's
$|\xi;k\ra$, eq.  (\ref{|xi;k>}), as well. The latter correspond to $u =
{\rm cosh}^2r,\quad v = {\rm sinh}^2r\,\exp(2i\theta),\quad w= {\rm
sinh}(2r)\,\exp(i\theta)$, where tanh$r=|\xi|$ and $\theta = {\rm arg}\xi
+\pi$. Moreover, if one calculate all the first and the second moments of
$K_i$ in $|\xi;k\ra$ \ci{T99b} one will find that they minimize
Schr\"odinger UR (\ref{xySUR}) for every pair $K_i$-$K_j$, i.e. the
standard $SU(1,1)$ CS's exhibit maximal uncertainty optimality. For the pair
$K_1$-$K_2$ this property of $|\xi;k\ra$ was discovered in \ci{T94}.
Furthermore the CS's $|\xi;k\ra$ are the unique states to exhibit this
maximal uncertainty optimality for the three observables $K_i$. This
unique property of $|\xi;k\ra$ can be proved most easily if one consider
the system of three eigenvalue equations (no summation over $i,\,j$)
\beq\lb{3eqs}         
(\beta_iK_i+\beta_jK_j)|\psi\ra = z_{ij}|\psi\ra,\quad i<j,
\eeq
every one of which is necessary and sufficient $|\psi\ra$ to minimize
(\ref{xySUR}) for $K_i$, $K_j$. In the standard $SU(1,1)$ CS's
representation \ci{ZhangAli} or in the BG analytic representation
(\ref{3eqs}) is a system of ordinary differential equations. In the BG CS
representation it is obeyed by the analytic function  $\exp(c
\eta)$ of $\eta$, $|c|\leq 1$, only, the latter corresponding to the CS's
$|\xi;k\ra$ (for details see the Appendix in \ci{T99}).  Let us recall
however that these group-related CS's can't exhibit squeezing in $\Dlt K_1$
and $\Dlt K_2$: $(\Dlt K_i)^2(\xi)\geq k/2$, $i=1,2$.

Similarly one can prove the maximal uncertainty optimality of the standard
$SU(2)$ group-related CS's. These results can be extended to semisimple Lie
groups -- the corresponding CS's with lowest/highest weight reference vector
are unique to minimize the Robertson UR for all generators and for the
quadratures of the Weyl ladder operators as well.

\section{\sm\bf CHARACTERISTIC UNCERTAINTY RELATIONS AND THEIR STATE 
EXTENSIONS}

From the matrix theory is known \ci{Gantm} that $\det M$ is an invariant
(under similarity transformations) characteristic coefficient of a matrix
$M$.  For an $n\times n$ matrix there are $n$ such invariant coefficients
$C_r^{(n)}$, $r= 1,2,\ld, n$, defined by means of the secular equation
\beq\lb{chareqn}   
0 = \det(M - \lam) = \sum_{r=0}^{n} C^{(n)}_r(M)(-\lam)^{n-r}.
\eeq
The characteristic coefficients $C^{(n)}_r$ are equal \ci{Gantm} to
the sum of all principal minors ${\cal
M}(i_1,\ldots,i_r;M)$ of order $r$. 
One has $C^{(n)}_0 = 1$, $C^{(n)}_1 = {\rm Tr}\,M = \sum m_{ii}$, and
$C^{(n)}_n = \det M$. For $n=3$ we have, for example, three principle
minors of order $2$.

In these notations Robertson inequality (\ref{RUR}) reads as 
$C^{(n)}_n\left(\sig(\vec{X})\right) \geq C^{(n)}_n\left(C(\vec{X})\right)$.
Inasmuch as the principal submatrices of order
$r$ of the dispersion matrix and of the mean commutator matrix  are in
fact again dispersion and mean commutator matrices for the $r$
observables, the Robertson UR was extended \ci{TD} in an invariant manner
to all characteristic coefficients in the form
\beq\lb{CUR}       
C^{(n)}_r[\sig(\vec{X})] \,\geq\, C^{(n)}_r[C(\vec{X})],
\eeq
$r=1,2,\ld,n$.
These invariant relations were called characteristic uncertainty relations
\ci{TD}. Robertson relation (\ref{RUR}) is one of them, and can be called
the $n$th-order characteristic inequality. Schr\"odinger UR (\ref{xySUR})
in the characteristic form reads as $C^{(2)}_2[\sig(X,Y)] \,\geq\,
C^{(2)}_2 [C(X,Y)]$.

The minimization of the first-order inequality in expression (\ref{CUR}),
Tr$\,\sig(\vec{X}) = {\rm Tr}\,C(\vec{X})$,  can occur in the case of
commuting operators only, as Tr$\,C(\vec{X})\equiv 0$.
An important example of minimization of the second-order inequality was
pointed out in Ref. 24. The spin and quasi-spin CS's $|\tau;j\ra$ and
$|\xi;k\ra$ minimize the second-order characteristic inequality for the
three generators $J_{1,2,3}$ and $K_{1,2,3}$  respectively. From the
results of Section 5 
(see also Ref. 47) it follows that the standard SU(1,1)
and SU(2) CS's are the unique states that minimize simultaneously the
second- and the third-order characteristic inequalities for the
corresponding three generators.

The characteristic inequalities relate combinations
$C_r^{(n)}[\sig(\vec{X};\rho)]$ 
of second moments of $X_1,\ld,X_n$ in a (generally mixed) state $\rho$ to
the combinations $C_r^{(n)}[C(\vec{X};\rho)]$ of first moments of their
commutators in the same state. However, there is no principal problem with
which to compare the statistical properties of observables in different
states.  From the mathematical point of view the derivation of the
state-extended characteristic UR's resorts on two simple matrix
properties:\,\, (a) The sum of symmetric (antisymmetric)
matrices is again a symmetric (antisymmetric) matrix and\, (b) the sum of
nonnegative-definite matrices is again a nonnegative-definite matrix. The
symmetricity of $\sig(\vec{X})$, the antisymmetricity of $C(\vec{X})$, and
the nonnegativity of $\sig(\vec{X})$ and of $R(\vec{X}) =
\sig(\vec{X})+iC(\vec{X})$  are the crucial properties, that the Robertson
derivation of the inequality (\ref{RUR}), and of (\ref{CUR}) as well,
relies on \ci{R34} (see also the proof in Ref. 47). Therefore we can
rewrite the characteristic UR's for the sum of several uncertainty and
mean commutator matrices that correspond to different, generally mixed,
states $\rho_m$, $m =1,\ld, m_s$,
\beq\lb{extendCUR} 
C_r^{(n)}\lf[{\textstyle\sum_m}\sig(\vec{X};\rho_m)\rt]
\,\geq\, C_r^{(n)}\lf[{\textstyle \sum_m}C(\vec{X};\rho_m)\rt].
\eeq
 These are extended characteristic uncertainty inequalities for $n$
observables (extended to several states).  For $r=n$ in inequality 
(\ref{extendCUR}) we have the extension of the Robertson relation to the
case of several states
\beq\lb{extendRUR} 
\det\lf[{\textstyle\sum_m}\sig(\vec{X},\rho_m)\rt] \,\geq\,
\det\lf[{\textstyle\sum_m}C(\vec{X},\rho_m)\rt].
\eeq
Inasmuch as $\det\sum \sig_m\neq \sum\det\sig_m$ these are indeed new
uncertainty inequalities, which extend the Robertson inequality to several
states. We note that extended relations (\ref{extendCUR}) and
(\ref{extendRUR}) are invariant under the nondegenerate linear
transformations of the operators $X_1,\ld,X_n$. If those operators span a
Lie algebra, then we obtain the invariance of inequality (\ref{extendCUR})
under the Lie group action in the algebra.  If in pure states $|\psi_m\ra$
inequality (\ref{extendRUR}) is minimized, then it is minimized also in
the states $U(g)|\psi_m\ra$ as well, where $U(g)$ is the unitary
representation 	of the group $G$. For two observables $X$ and $Y$ and two
states $|\psi_{1,2}\ra$ that minimize Schr\"odinger inequality
(\ref{xySUR}), inequality (\ref{extendRUR}) produces
\bear\lb{extendSUR} 
\frac 12 \left[\Dlt XX(\psi_1)\Dlt YY(\psi_2) +
\Dlt XX(\psi_2)\Dlt YY(\psi_1)\right]\nn \\
 - \Dlt XY(\psi_1)\Dlt XY(\psi_2)\nn \\ 
\,\geq\, \frac 14\la\psi_1|[X,Y]|\psi_1\ra \la\psi_2|[Y,X]|\psi_2\ra,
\eear
where, for the sake of symmetry, the variance $(\Dlt X)^2$ of $X$ in
$|\psi\ra$ was denoted as $\Dlt XX(\psi)$. One can prove that this UR
remains valid for any state \ci{T'00}. It can be considered 
one of the basic UR's for quantum states. One can see that, if the two
states coincide, inequality (\ref{extendSUR}) recovers inequality
(\ref{xySUR}). The minimization properties of the extended UR's remain to
be considered elsewhere. Here we note that for $X=q$ and $Y=p$, inequality
(\ref{extendSUR}) is saturated by any two Fock states, Glauber CS's, or
both, for example. The relation is also minimized in two equally  squeezed
states $|\alf_1,u,v\ra$ and $|\alf_2,u,v\ra$, Im$(uv^*)=0$.

It is worth noting that, at $X=Y$, UR (\ref{extendSUR}), and
(\ref{xySUR}) as well, does not survive. This result encouraged me 
to look for an UR for two states and one observable. One solution is ($\la
i|X|j\ra = \la\psi_i|X|\psi_j\ra$)
\begin{eqnarray}\lb{extendSUR2}            
\hs{-5mm}\left[(\Dlt X(\psi_1))^2+\la1|X|1\ra^2\right]
\left[(\Dlt X(\psi_2))^2+ \la2|X|2\ra^2\right]\nn \\
\,\geq\, \lf|\la1|X^2|2\ra\rt|^2.
\end{eqnarray}
Both inequalities (\ref{xySUR}) and (\ref{extendSUR2}) follow from the
Schwarz inequality, $|\la\Psi_1|\Psi_2|^2 \leq ||\Psi_1||^2||\Psi_2||^2$,
with suitably chosen $|\Psi_1\ra$ and $|\Psi_2\ra$. Inequality
(\ref{extendSUR}) is different.

The extended UR can be used for construction of distances between
quantum states. One simple new distance is based on UR
(\ref{extendSUR2}), $D^2[\psi_1,\psi_2] = 2\left(1 -
g(\psi_1,\psi_2;X)\right)$, where
\beq\lb{g[]}        
g(\psi_1,\psi_2;X) =   \frac{\left|\la\psi_2|X^2|\psi_1\ra\right|}
{\left(\la\psi_1|X^2|\psi_1\ra\la\psi_2|X^2|\psi_2\ra\right)^{1/2}},
\eeq
and $X$ is any continuous or strictly positive observable. In this case
$g(\psi_1,\psi_2;X) = 1$ if and only if $|\psi_1\ra=|\psi_2\ra$, and 
$D^2[\psi_1,\psi_2]$ does satisfy all the requirements for a distance. In
particular one can take $X=1$, which reproduces the well
known  Bures--Uhlmann distance (see the references in Ref. 60). In this
way we establish the relation between the extended UR and the polarized
distances \ci{distance}.

Finally it is worth noting that every extended characteristic inequality
can be written in terms of two new positive quantities, the
sum of which is not greater than unity.  To this end 
we put $C_r^{(n)}[\sig(\vec{X},\rho)] = \alf_r (1-P_r^2)$,
where $0\leq P_r^2 \leq 1$ (i.e., $1-P_r^2 \leq 1$) and $\alf_r \neq
0$. For $r=n$ one has (omitting index $r=n$)
det$\sig(\vec{X},\rho) = \alf\,(1-P^2)$. $\alf_r$ may be viewed as scaling
parameters. Then we can set $C_r^{(n)}[C(\vec{X},\rho)] = \alf_r V_r^2$ and
obtain from inequality (\ref{CUR}) the inequalities for $P_r$ and $V_r$,
r=1,\ld,n, 
\beq\lb{CUR2} 
P_r^2(\vec{X},\rho) + V_r^2(\vec{X},\rho) \leq 1.
\eeq
The equality in expression (\ref{CUR2}) corresponds to the equality in
expression (\ref{CUR}) or (\ref{extendCUR}).  $P_r,\,V_r$ are functionals
of the state $\rho$ [or of $\rho_1,\rho_2,\ld$ in the case of extended
inequalities (\ref{extendCUR})]. These can be called complementary
quantities, and the form (\ref{CUR2}) of the extended characteristic
relations can be called complementary form. Let us note that $P_r$ and
$V_r$ are not uniquely defined by  $\sig$ and $C$; they  depend on the
choice of the scaling parameter $\alf_r$. In the case of bounded operators
$X_i$ (say, spin components) the characteristic coefficients of $\sig$ and
$C$ are also bounded.  Then $\alf_r$ can be taken as the inverse maximal
value of $C_r^{(n)}(\sig)$. For one state and two observables with only
two eigenvalues each, the complementary inequality (\ref{CUR2}) was
recently considered in the important paper by Bj\"ork {\it et al}
\ci{Bjork}.  In this case the meaning of the complementary quantities $P$
and $V$ was elucidated to be that of the predictability ($P$) and the
visibility ($V$) in the {\it welcher weg} experiment \ci{Bjork}.

It is worth underlining that we have considered the developments of
generalization of the SS's and UR's mainly along the lines of
characteristic invariants of the uncertainty matrix. Schr\"odinger UR
(\ref{pqSUR}) is the simplest ordinary characteristic UR. Other types of
ordinary UR's are also discussed in the literature \ci{SCB95,T97}\,$^,$
\ci{DM87}$^-$\ci{others}; UR for higher moments and universal invariants
\ci{DM82}, trace-class UR's \ci{SCB95,T97}, parameter-based UR's \ci{BC},
minimal-length UR's \ci{KMM},  etc.\,\,  For an exhaustive list of
references through 1986 see the review in Ref. 61.

\section{\sm\bf CONCLUSION}

  The set of the characteristic uncertainty relations (UR's) and the
related squeezed states (SS's) are briefly reviewed and compared in
accordance with the generalizations of the three equivalent definitions of
the canonical coherent states (CS's). It was shown that the multimode
canonical SS's are the unique states  (so far) for which the three
definitions are equivalently generalized, where the basic uncertainty
relation being that of Robertson (\ref{RUR}).  It was noted that the
group-related CS's with the lowest (highest) weight reference vector
minimize the Robertson relation for all generators and for the quadratures
components of the Weyl ladder operators as well.  The minimization of the
other characteristic inequalities [inequality (\ref{CUR})] can be used for
finer classification of group-related CS's.  The standard SU(1,1) CS's
were shown to be the unique states that minimize the third- and the
second-order characteristic inequalities for the three generators.  For
two observables a new inequality, less precise than that of Heisenberg, is
described that is minimized in Barut-Girardello-type CS's only.

It was proved that the characteristic uncertainty inequalities can be
naturally extended to the case of several states.  The state-extended
uncertainty inequalities  can be used for the construction of distances
between quantum states. Further properties and applications of the new
uncertainty relations remain to be considered elsewhere.

It was also shown here that the characteristic inequalities can be written
in complementary form  in terms of two positive quantities less than
unity. In the case of one state and two observables with two eigenvalues
each, the meaning of these complementary quantities were recently
elucidated \ci{Bjork} to be that of the predictability and visibility in
the {\it welcher weg} experiment. \\[-5mm]

\subsection*{Acknowledgment}
\vspace{-5mm} The author is grateful to V.I. Man'ko and to a second referee
for valuable remarks.\\

The author's e-mail is dtrif@inrne.bas.bg.
\newpage
\vs{45mm}

{\sm \bf REFERENCES} 

\end{document}